\documentclass[twocolumn,trackchanges,times]{aastex62}
\def\mathbi#1{\textbf{\em #1}}
\newcommand{\sm}{$\sim$}

\newcommand{\hmi}{Helioseismic and Magnetic Imager}
\newcommand{\aia}{Atmospheric Imaging Assembly}
\newcommand{\Sdo}{\textit{Solar Dynamics Observatory}}
\newcommand{\sdo}{\textit{SDO}}
\newcommand{\GOES}{\textit{Geostationary Operational Environmental Satellite}}

\newcommand{\kms}{km~s$^{-1}$}
\newcommand{\Hsi}{\textit{Reuven Ramaty High Energy Solar Spectroscopic Imager}}
\newcommand{\hsi}{\textit{RHESSI}}
\newcommand{\trace}{\textit{TRACE}}
\usepackage{amsmath}
\usepackage{color,colordvi}

\turnoffediting

\begin{document}

\title{AN ERUPTIVE CIRCULAR-RIBBON FLARE WITH EXTENDED REMOTE BRIGHTENINGS}

\author{Chang Liu}
\affil{Institute for Space Weather Sciences, New Jersey Institute of Technology, University Heights, Newark, NJ 07102, USA; \href{mailto:chang.liu@njit.edu}{chang.liu@njit.edu}}
\affil{Big Bear Solar Observatory, New Jersey Institute of Technology, 40386 North Shore Lane, Big Bear City, CA 92314, USA}
\affil{Center for Solar-Terrestrial Research, New Jersey Institute of
Technology, University Heights, Newark, NJ 07102, USA}

\author{Avijeet Prasad}
\affil{Center for Space Plasma and Aeronomic Research, The University of Alabama in Huntsville, Huntsville, AL 35899, USA}

\author{Jeongwoo Lee}
\affil{Institute for Space Weather Sciences, New Jersey Institute of Technology, University Heights, Newark, NJ 07102, USA; \href{mailto:chang.liu@njit.edu}{chang.liu@njit.edu}}
\affil{Big Bear Solar Observatory, New Jersey Institute of Technology, 40386 North Shore Lane, Big Bear City, CA 92314, USA}
\affil{Center for Solar-Terrestrial Research, New Jersey Institute of
Technology, University Heights, Newark, NJ 07102, USA}

\author{Haimin Wang}
\affil{Institute for Space Weather Sciences, New Jersey Institute of Technology, University Heights, Newark, NJ 07102, USA; \href{mailto:chang.liu@njit.edu}{chang.liu@njit.edu}}
\affil{Big Bear Solar Observatory, New Jersey Institute of Technology, 40386 North Shore Lane, Big Bear City, CA 92314, USA}
\affil{Center for Solar-Terrestrial Research, New Jersey Institute of
Technology, University Heights, Newark, NJ 07102, USA}

\begin{abstract}
We study an eruptive X1.1 circular-ribbon flare on 2013 November 10, \edit2{combining multiwavelength observations with a coronal field reconstruction using a non-force-free field method}. In the first stage, a filament forms via magnetic reconnection between two mildly twisted sheared arcades, \edit1{which are embedded under the fan dome associated with a null point}. This reconnection seems to be driven by photospheric shearing and converging flows around the inner two arcade footpoints, \edit2{consistent with } the flare-related changes of \edit2{transverse } field. The southern portion of the filament rises upward due to torus instability and pushes against the null point. The induced null point reconnection then generates the circular ribbon and the initial remote brightening in the west, as accelerated electrons precipitate along the fan and propagate outward \edit2{along quasi-separatix surfaces with high values of the squashing factor (Q) } \edit1{in the envelope fields, which have a curtain-like shape here}. In the second stage, the southern end of the flux rope breaks away from the surface, sequentially disrupts the dome and overlying fields, and erupts in a whipping-like fashion to become a partial halo coronal mass ejection. This leads to an enhanced flare emission and fast-moving remote brightenings at the footpoints of the magnetic curtain, which span a remarkably broad region and are also associated with coronal dimmings. This is a rare example of eruptive circular-ribbon flares, in which the evolution of a flux rope from its formation to successful eruption out of the dome and the resulting unusually extended remote brightenings are completely observed.

\end{abstract}


\section{INTRODUCTION}\label{introduction}
Solar flares and coronal mass ejections (CMEs) are believed to be powered by magnetic energy stored and accumulated in the corona and abruptly released by magnetic reconnection. The accelerated particles precipitate along the reconnecting field lines in separatrix or quasi-separatrix layer (QSL) surfaces (dividing domains with discontinuous or strong gradients of field line connectivity), and subsequently produce bright flare ribbons in the lower chromosphere \citep[e.g.,][]{demoulin96,janvier13}. Hence, the morphology and dynamics of chromospheric ribbons can be a robust mapping of the structure and evolution of coronal reconnection region. Observations thus far have revealed that flare ribbons can assume various geometric shapes, from simply elongated to more complex J-shaped, X-shaped, and circular ribbons \citep[see e.g.,][and references therein]{toriumi19}. This implies that beyond the standard two-dimensional model \citep[e.g.,][]{kopp76}, three-dimensional (3D) flare models are often needed to depict the topology of flaring magnetic fields \citep[e.g.,][]{galsgaard97,galsgaard03}.

Notably, circular-ribbon flares have been studied extensively using observations from the \Sdo\ \citep[\sdo;][]{pesnell12} (see \citealt{liuc19} and references therein), after they were \edit1{reported} using \trace\ images \citep{masson09} and later explored using the digitized solar film data \citep{wang12}. This type of flares is deemed important as the appearance of a circular ribbon is indicative of 3D magnetic reconnection around a null point (NP) \citep{lau90,torok09,pontin13}, which represents a magnetic singularity. Specifically, in the fan-spine-NP topology, the spine field lines inside/outside the dome-shaped fan meet at the NP. If reconnection occurs at the NP, the circular ribbon (the accompanied inner/outer compact ribbons) is expected to form at the intersection of the fan surface (inner/outer spine field lines) with the chromosphere. Often a sequential brightening of the circular ribbon and elongated spine-related ribbons are observed, suggesting that the fan and spine could be embedded in extended QSLs \citep{masson09,reid12}. This can be visualized by rendering 3D views of squashing factor $Q$ \citep[e.g.,][]{yang15,zhong19}, which measures the gradient of field line connectivity and hence can gauge the strength of QSLs \citep{titov02,titov07}.

As shown in a majority of previous observational studies, circular-ribbon flares are usually confined events, while some of them can occur in tandem with a CME \citep{song18}. These eruptive events are particularly interesting because they involve the interaction between eruptive features with the fan-spine fields that have a confined nature \citep{lee16a}. Largely the following three scenarios have been suggested based on limited samples. (1) The NP reconnection weakens the overlying field, leading to the eruption of a filament flux rope (FR) initially lying under the fan dome \citep{sun13}. (2) Such a filament may erupt first due to instability and subsequently triggers the NP reconnection \citep{jiang14,zhang15,joshi15,liu15,lee16a}. (3) The outer spine-related fields may comprise a twisted FR, which erupts and then causes reconnection at the null \citep{liuc19}. Obviously, more detailed case analyses of circular-ribbon flares, especially eruptive ones, are desired, as studying this particular type of flares can help understand solar eruptions in general \citep{masson17}.

In this work, we present a GOES-class X1.1 circular-ribbon flare on 2013 November 10 in NOAA active region (AR) 11890, which is associated with a partial halo CME. \edit1{In the above studies of CME-associated circular-ribbon flares}, the filament under the dome is already well formed right before the flare. Here, we witness the formation of an eruptive \edit2{filament or FR } from interaction of two sheared arcade (SA) systems and its ensued whipping-like asymmetric eruption throughout the dome. \edit1{As previously reported, filaments of different sizes can be formed with the presence of separatrices or QSLs under dome-like structures before eruptions \citep[][and references therein]{chandra17,fuentes18}}. In this event the successful eruption clearly excites a rapidly evolving, elongated remote brightening region, which is significantly much extended than the outer-spine-related ribbon in other reported circular-ribbon flares. \edit1{Our main objective is to explain the above distinct observations with aid of a preflare coronal field model reconstructed with a non-force-free field (NFFF) approach and photospheric vector field data. Compared to other extrapolation methods, the NFFF reflects a more realistic boundary condition.} This paper is planned as follows. In Section~\ref{data}, we introduce observational data and analysis methods, including flow tracking. Methods of the NFFF extrapolation and 3D magnetic field analysis are described in Section~\ref{nfff} \edit1{and the Appendix}. In Section~\ref{topology}, the preflare magnetic field structure is elucidated based on observations and the field extrapolation model. With the 3D field configuration in mind, we analyze the event progression in Section~\ref{evolution} and summarize and discuss major findings in Section~\ref{summary}.

\section{OBSERVATIONS AND DATA ANALYSIS}\label{data}
This impulsive flare has a short duration in 1--8~\AA\ soft X-ray (SXR) flux as recorded by the \GOES, with the start/peak/end times at 05:08/05:14/05:18~UT on 2013 November 10. We used hard X-ray (HXR) light curves registered by the \Hsi~\citep[\hsi;][]{lin02} to help differentiate different stages of event temporal evolution. To observe flare ribbons and follow large-scale eruptive activities, we utilized 304~\AA\ (He~{\sc ii}; 0.05~MK), 1600~\AA\ (C~{\sc iv}; 0.1~MK), and 211~\AA\ (Fe~{\sc xiv}; 2.0~MK) images taken by the \aia\ \citep[AIA;][]{lemen12} on board \sdo.

Full-disk photospheric vector magnetograms at 1\arcsec\ resolution are available from \sdo's \hmi\ \citep[HMI;][]{schou12}. These data products are produced by the HMI data processing pipeline, including Stokes inversion of the Fe~{\sc i} 617.3 nm absorption line \citep{borrero11}, disambiguation of azimuthal angle \citep{hoeksema14}, etc. The precision of the line-of-sight and \edit2{transverse} fields is on the order of 10 and 100~G, respectively. We further derotated and remapped the retrieved vector data with a Lambert (cylindrical equal area) projection (also applied to the AIA data for comparison), using Solar SoftWare procedures provided by the HMI team. To investigate the surface magnetic and flow fields evolution, we chose the 135~s cadence HMI data series \citep{sun17}. Photospheric flow fields were derived using the differential affine velocity estimator for vector magnetograms \citep[DAVE4VM;][]{schuck08} method. The feature tracking adopted a window size of 19 pixels following the suggestion from previous studies \citep[e.g.,][]{liuy13}.

\section{ANALYSIS OF MAGNETIC FIELDS}\label{nfff}
In an effort to reveal the 3D magnetic field structure and topology, we carried out a numerical NFFF extrapolation by using the code developed by \citet{hu08a} and \citet{hu08b,hu10}. \edit1{A brief discussion of this technique is provided in the Appendix.} NFFFs constructed this way were used as the initial condition for magnetohydrodynamic simulations, the results of which showed a high similarity to observations \citep{prasad17,prasad18,prasad19,nayak19}. \edit1{For the} NFFF modeling of the present AR, we picked a set of vector magnetograms at a preflare time from the less noisier, 12 minute cadence HMI data series, and cut a large area of 880~$\times$~1300 pixels$^2$ as the bottom boundary. This encloses the entire circular-ribbon and extended remote brightening regions. The height of the computation box was set to 880 pixels.

In the modeled 3D magnetic field, we searched for coronal NPs using a code based on the Poincar{\'e}-Hopf theorem \citep{zhao05}. We traced magnetic field lines to calculate the squashing factor $Q$ and the magnetic twist number $\mathcal{T}_w$ using the codes developed by \citet{liur16}. Regions with $Q \gg 1$ can be regarded as QSLs \citep{titov02}. The twist number of a field line that characterizes its winding is given by $\mathcal{T}_w=\int_L(\nabla\times\mathbi{B})\cdot\mathbi{B} / (4\pi B^2)\,dl$, where $L$ is the length of field line \citep{berger06}. $\mathcal{T}_w = 1$ is usually used to discriminate SAs from twisted FRs \citep[e.g.,][]{wang15b,liur16}. Also, a FR may erupt due to torus instability, when its axis reaches the height $h$ where the decay index $n = -d {\rm log}(B)/d {\rm log}(h)$ of the strapping field $B$ surpasses the torus unstable threshold of 1.5 \citep[e.g.,][]{bateman78,kliem06,torok07,liuy08,demoulin10}.

\begin{figure}[!t]
\epsscale{1.17}
\plotone{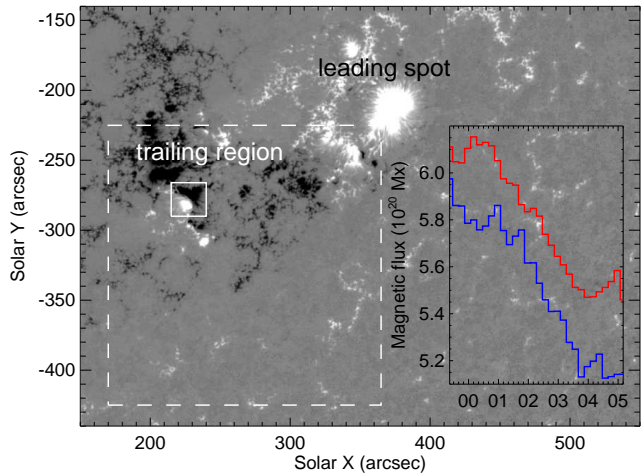}
\caption{HMI $B_z$ image of NOAA AR 11890 at 04:46:12~UT on 2013 November 10. The inset shows the evolution of positive (red) and absolute negative (blue) magnetic flux within the small $\delta$-spot region (solid box) calculated with 12-minute HMI data, from 2013 November 9 23:22:12 to November 10 05:10:12~UT. The abscissa is time in units of hours. \edit2{The $\delta$ spot has the negative flux much larger than the positive one. To display both fluxes in one panel, we reduce the absolute negative flux by subtracting 7~$\times$~10$^{20}$~Mx from it}. The dashed box represents the field of view of Figures~\ref{f2}(a) and (b). \label{f1}}
\end{figure}
\begin{figure*}[!h]
\epsscale{1.17}
\plotone{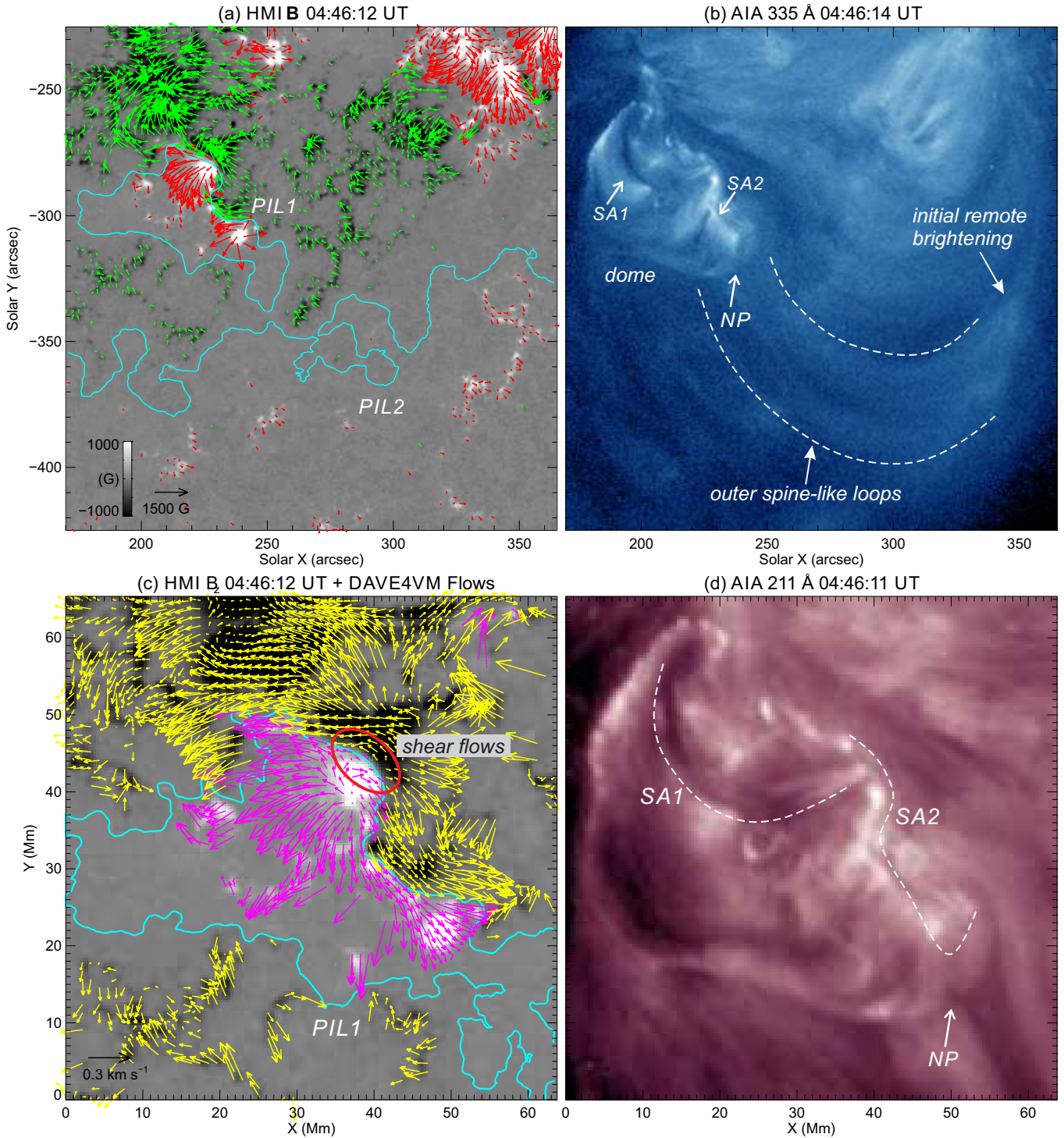}
\caption{Overall preflare structure of active region. (a) HMI $B_z$ overplotted with arrows representing $B_t$ vectors (only plotted for locations with $|B_z| \geqslant 100$~G). For clearness, arrows in positive (negative) magnetic polarity are coded in red (green). The cyan lines are the main PIL1 and the secondary PIL2 (smoothed by a window of 9\arcsec$\times$~9\arcsec). (b) AIA 335~\AA\ image showing the outer spine-like loops (\edit2{lying} approximately between the dashed lines) that extend from the dome to the initial remote brightening region. The position of the coronal NP found in the NFFF model (see Figure~\ref{f3}) is indicated by an arrow. SA1 and SA2 are also marked. (c) HMI $B_z$ in the flare core region superimposed with arrows representing photospheric \edit2{transverse} flow vectors tracked with DAVE4VM (averaged between 04:30:27 and 05:01:07~UT). Obvious shearing/converging flows along the PIL1 (smoothed by a window of 2.5\arcsec$\times$~2.5\arcsec) can be observed \edit2{approximately within the red ellipse}. (d) AIA 211~\AA\ image of the circular-ribbon flaring region (same as the field of view of (c)), with the SA1 and SA2 delineated by dashed lines. The position of NP is also indicated by an arrow. \label{f2}} 
\end{figure*}
\begin{figure*}
\epsscale{1.09}
\plotone{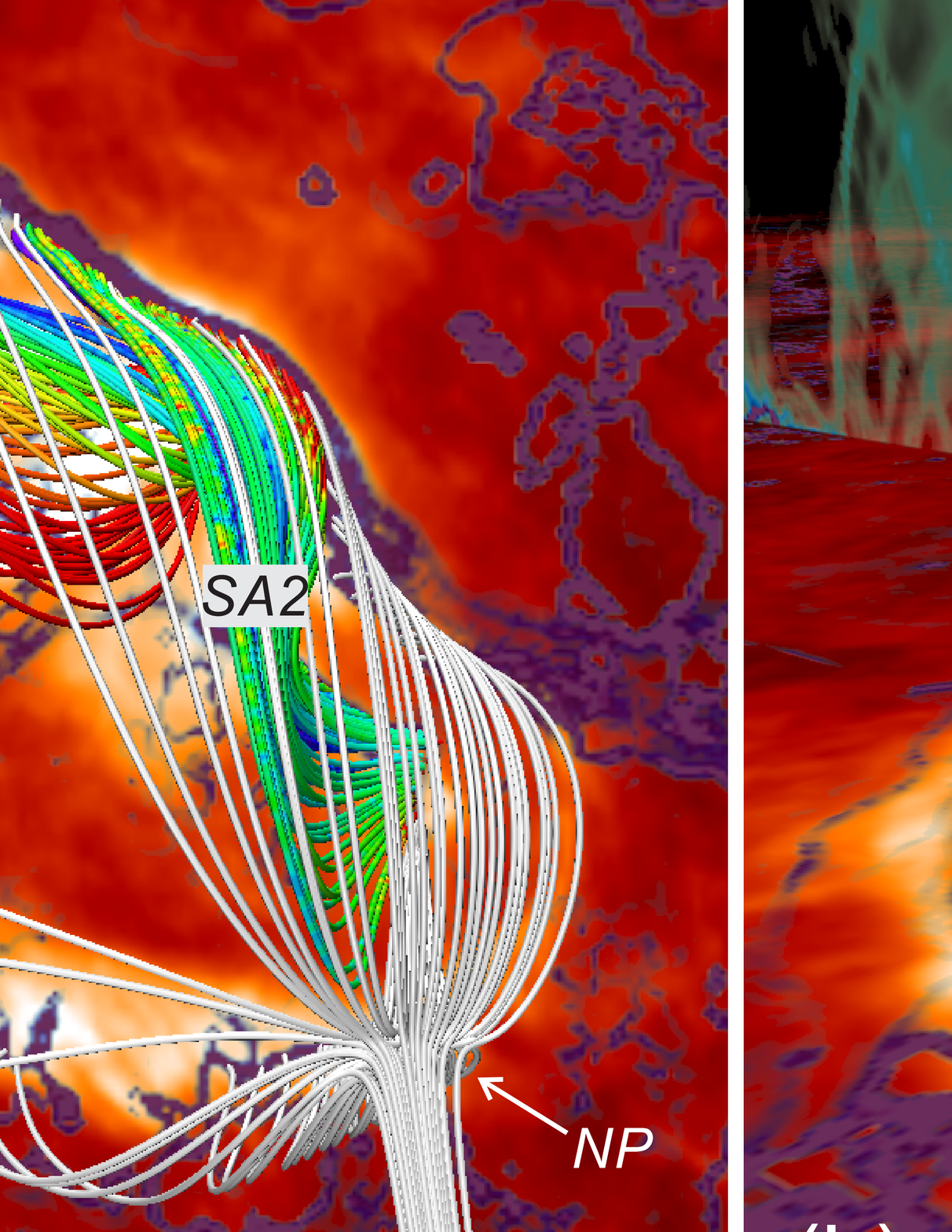}
\caption{Preflare NFFF model at 04:46:12~UT, in top (a and d) and perspective (b--c and e) views. In (a)--(c), the field lines color-coded based on $\mathcal{T}_w$ illustrate the SA1 and SA2 along the main PIL1. The gray lines are selected field lines that approach the NP, \edit1{illustrating the fan-spine structure}. The background displays the AIA 304~\AA\ image of the circular ribbon at 05:14:10~UT around the SXR peak \edit2{together} with log$Q$ computed with the NFFF model at $z=0$ (photosphere). The high-$Q$ region on the surface is seen in dark purple. Also shown in (b) is log$Q$ in a vertical cutting plane passing through the NP and the footpoint of inner spine, and in (c) is the decay index in another vertical cutting plane passing through the NP and the northern footpoints of SA1. Drawn in (d) and (e) are the same field lines and background as plotted in (a)--(c), together with curtain-like cyan field lines that overlie the dome and connect to the extended remote brightening regions. Structures in dark purple are the 3D volume rendering of high-$Q$ surfaces. \label{f3}} 
\end{figure*}

\section{PREFLARE MAGNETIC FIELD STRUCTURE}\label{topology}
Although the source NOAA AR 11890 with a main leading spot does not show clear signs of overall nonpotentiality, the SOL2013-11-10T05:14 X1.1 flare occurs upon polarity intrusion into the trailing region \citep{schrijver16}, \edit1{as shown in Figure~\ref{f1}. The opposite polarities of the small $\delta$-spot region (solid box) shows continuous flux cancellation from about 01--04~UT (see the inset) and persistent shearing and converging motions till nearing the flare time (see below). Figures~\ref{f2}(a) and (b) display} the photospheric and coronal environment \edit1{of a subregion of this AR (indicated by the dashed box in Figure~\ref{f1})} at 04:46~UT (about 22 minutes before the flare start), \edit1{focused on the present circular-ribbon flare}. As seen in Figure~\ref{f2}(a), positive polarities are surrounded by the trailing negative fields, forming a quasi-circular magnetic polarity inversion line (PIL; here labeled as PIL1). Figure~\ref{f2}(c) gives a zoomed-in view of the flare core region overplotted with flow field vectors derived with DAVE4VM. One can clearly see that in the \edit1{$\delta$-spot region} along the PIL1 (enclosed by the red ellipse), there exists pronounced shearing (i.e., oppositely directed flows on the two sides of a PIL) and converging flows \edit1{just preceding the flare start time}. These flows are known to be closely associated with flare triggering \citep[e.g.,][]{yang04,deng06,deng11,welsch09,wang15b}. In the AIA 211~\AA\ images, a pair of arcade loops SA1 and SA2 \edit2{lie close to} each other and their inner footpoints are rooted in the shearing/converging flow region (see Figure~\ref{f2}(d)). Both SA1 and SA2 most probably harbor filament materials, and appear to lie beneath a dome-like structure in 335~\AA\ EUV channel (see Figure~\ref{f2}(b)); also, large-scale outer spine-like loops apparently extend from the southwest portion of the dome and connect to a western remote region of positive magnetic polarity (where the initial remote brightening occurs; see Section~\ref{evolution} for details), overarching a secondary PIL2 running approximately southeast to northwest (Figure~\ref{f2}(a)). Interestingly, at the connecting interface between the dome and outer spine-like loops, a seemingly void region is enveloped by adjacent loops, mimicking a coronal NP (pointed to by the arrow in Figures~\ref{f2}(b) and (d)). Therefore, all the key magnetic topological features of a classical circular-ribbon flares are present in this flaring region.

To corroborate the flaring magnetic structure inferred from observations, we explore the 3D NFFF data cube constructed at the same preflare time. In Figures~\ref{f3}(a) (top view) and \ref{f3}(b)--(c) (perspective view), we trace magnetic field lines (gray) adjacent to a coronal NP, which we find to have a height of $z=14.3$~Mm and is located at the expected position in the southwest portion of the flare core region. We also trace two magnetic field bundles (SA1 and SA2) from the initial four footpoint-like brightening regions at the event onset (see Section~\ref{evolution}) and color code them based on $\mathcal{T}_w$. Also plotted in Figure~\ref{f3}(b) is log$Q$ in a vertical cutting plane passing through the NP and the footpoint of inner spine, and in Figure~\ref{f3}(c) is the decay index $n$ in another vertical cutting plane passing through the NP and the northern footpoints of SA1. Figures~\ref{f3}(d) (top view) and \ref{f3}(e) (perspective view) portray the magnetic field structure in a larger scale, in which we further reveal outer spine-like field lines (cyan) that envelop the flare core. The superimposed structures in dark purple are the volume rendering of high-log$Q$ surfaces in 3D. In all the panels, the image plotted as the background is a blend of 304~\AA\ image around the SXR peak (at 05:14:10~UT) with the log$Q$ map computed at $z=0$ (photosphere).

Based on these results, we see that the modeled dome-shaped fan structure, inner/outer spines, and the two mildly twisted SA systems embedded underneath closely resemble the EUV observation of the flare core (cf. Figure~\ref{f2}(b)). Compared with SA1, SA2 (especially its southern cusp-like portion close to the NP) could more readily be subject to the torus instability, as a result of the asymmetric strapping fields overlying SA1 and SA2. Near the inner two footpoints of SA1/SA2, there exists shearing/converging flows as observed. In large scale, we note that (1) the outer spine extends southward and connects to one portion of the remote brightenings (Figure~\ref{f3}(d)). (2) The overall outer spine-like field lines form a ``magnetic curtain'' that hangs above the extended arc-like remote brightening regions in positive magnetic fields (cf. Figures~\ref{f2}(a) and (b); see Section~\ref{evolution}) and converges towards the dome. Intriguingly, the portion of the curtain with the highest $Q$ mimics the mildly hot loops seen in 335~\AA, spreading from the dome to a western region where the remote brightenings are firstly excited (cf. Figures~\ref{f2}(b) and \ref{f3}(d)). In general, the fan and spine-related fields are rooted in high-$Q$ structures that outline the footprints of prominent QSLs, which are generally well cospatial with the observed flare ribbons and remote brightenings.

\begin{figure}
\epsscale{1.17}
\plotone{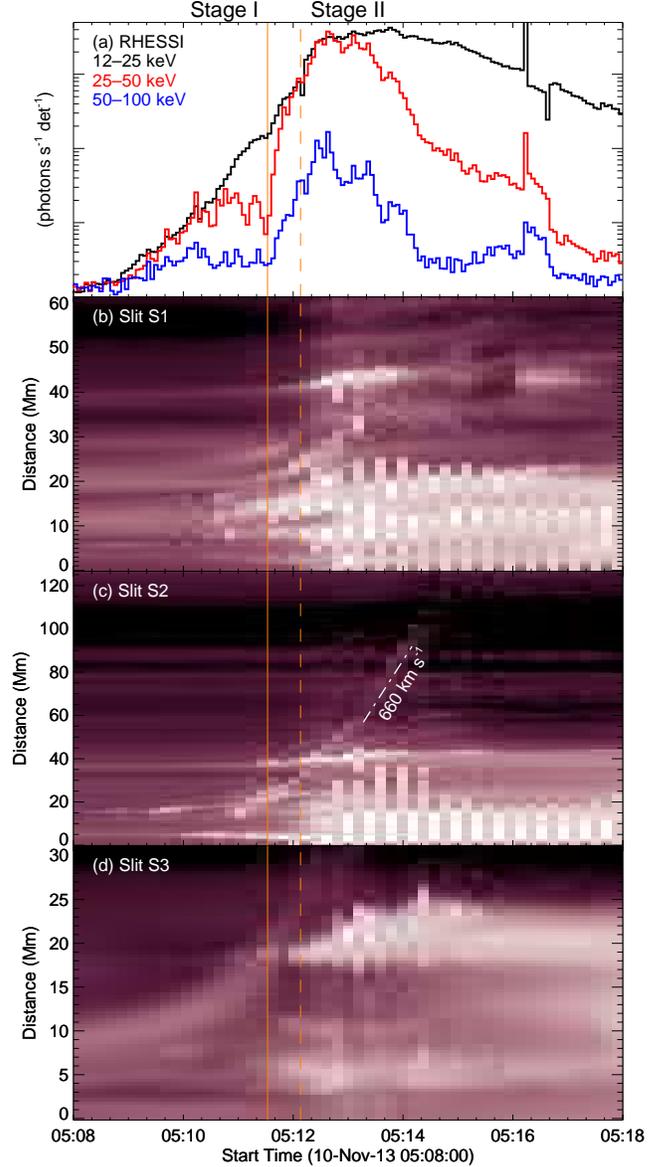}
\caption{Event temporal evolution. (a) Time profiles of \hsi\ photon rates (on arbitrary logarithmic scales), showing two stages (I and II) of event evolution with the second stage starting at \sm05:11:32~UT (indicated by the vertical solid line). Spurious jumps at 05:16:12 and 05:16:40~UT are caused by attenuator changes. (b)--(d) \edit2{Time-distance plots} for the slits S1, S2, and S3 marked in Figure~\ref{f5}. The distance is measured from the northern ends of the slits. The southern footpoints of the erupting loops (see Figure~\ref{f5}) detach from the surface around 05:12:08~UT, as denoted by the vertical dashed line. \label{f4}} 
\end{figure}
\section{EVENT EVOLUTION}\label{evolution}
\begin{figure*}
\epsscale{1.17}
\plotone{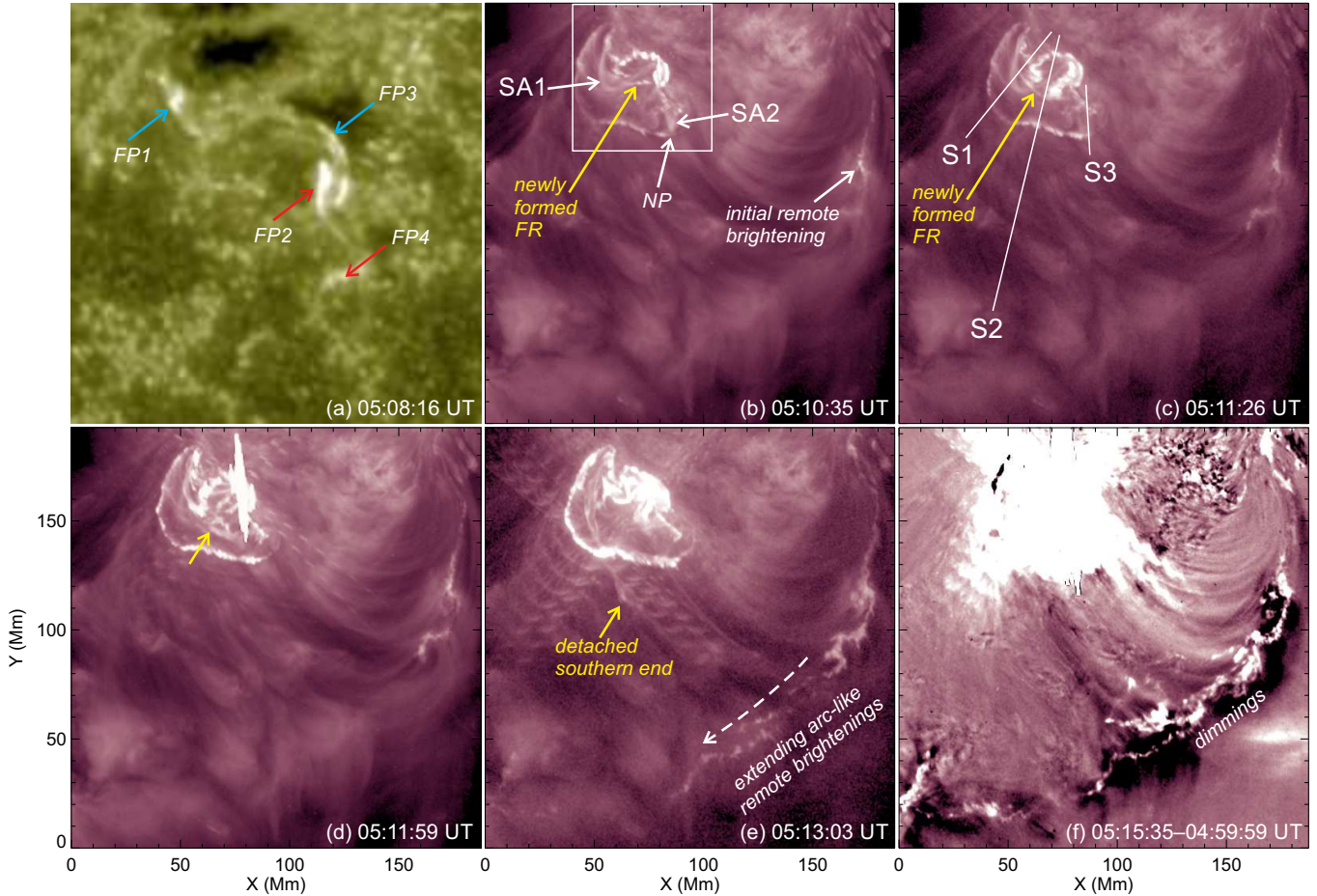}
\caption{Evolution of circular-ribbon flare region. (a) AIA 1600~\AA\ image showing largely four footpoint-like brightening regions FP1--FP4 at the event onset, with FP2/FP4 (FP1/FP3) located in the positive (negative) magnetic polarity field as pointed to by the red (cyan) arrows. (b)--(e) Selected AIA 211~\AA\ images showing the occurrence sequence of event key features, including the rising SA1 and SA2, the initial remote brightening region, the newly formed erupting FR, the detachment of their southern footpoints from the surface, and the moving remote brightening regions that rapidly extend southeastward. The box in (b) denotes the field of view of (a). \edit2{Time-distance plots for} the slits S1--S3 drawn in (c) are displayed in Figures~\ref{f4}(b)--(d). (f) AIA 211~\AA\ difference image showing coronal dimmings at locations cospatial with the remote brightenings. An animation is available, with the left animation panel showing (b)--(e) images of 211~\AA\ and the right animation panel showing (f) \edit2{base} difference images (relative to 04:59:59~UT). These sequences start at 2013 November 10 04:59:59~UT and end at 05:29:59~UT. The video duration is 10~s. \\(An animation of this figure is available.)\label{f5}} 
\end{figure*}

\begin{figure*}
\epsscale{1.17}
\plotone{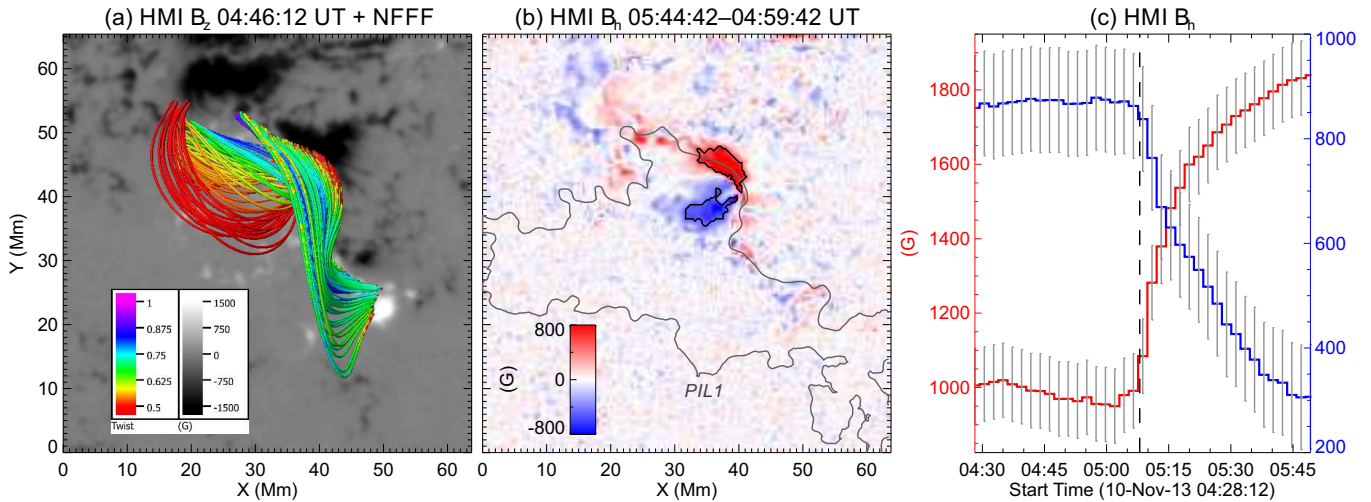}
\caption{Field structure and postflare changes. (a) HMI $B_z$ overplotted with same NFFF lines of SA1 and SA2 as in Figure~\ref{f3}(a). (b) Difference image of HMI $B_t$ between the pre- and postflare states, showing regions of the enhanced (red) and weakened (blue) $B_t$. The levels of black contours are 500 and $-$450~G. The gray contour is the PIL1, same as that shown in Figure~\ref{f2}(c). (c) Red (blue) lines show the temporal evolution of enhanced (weakened) $B_t$ averaged within the contoured regions in (b). Error bars represent an uncertainty of 100~G in the HMI \edit2{transverse} field data. The vertical dashed line denotes 05:08~UT, the flare start time in 1--8~\AA\ SXR. \label{f6}} \end{figure*}

In this section, we describe the event evolution in detail using multiwavelength observations. In Figure~\ref{f4}, the event temporal evolution is reflected by \hsi\ HXR light curves in several energy bands, which show \edit2{a series of minor peaks} followed by a major one, implying two distinct evolution stages (see more detailed description below). We attempt to link this to the dynamics of erupting flux loops as depicted by the time-distance profiles \edit2{drawn for} slits S1--S3 \edit2{across different event structures}. We show time sequence of EUV images in Figure~\ref{f5} and the supplementary movie, and incorporate changes of photospheric vector field from the pre- to postflare states in Figure~\ref{f6}. We mainly concern ourselves with understanding the mechanisms of event triggering and progression, by synthesizing observations of various event features with the magnetic structure and topology that we learned in Section~\ref{topology}.

Stage I (the first row of Figure~\ref{f5}; before \sm05:11:32~UT, the vertical solid line in Figure~\ref{f4}): It is spotted that at the earliest event onset around 05:08~UT, four compact brightenings FP1--FP4 along the PIL1 appear at the footpoints of SA1 and SA2 (Figures~\ref{f5}(a) and (b)). Soon afterward (from \sm05:09~UT), SA1 and SA2 seemingly start to grow (presumably rise upward), as can be seen in the time-distance diagrams for the slits S1 and S3 in Figures~\ref{f4}(b) and (d), respectively. From \sm05:09:30~UT, we observe the following two remarkable development. First, loops that are obviously newly formed (pointed to by the yellow arrow in Figures~\ref{f5}(b) and (c)) begin to brighten and gradually erupt outward, as shown in the time slice for the slit S2 in Figure~\ref{f4}(c). Second, at this time the southern cusp-like portion of SA2 seems to reach the NP region (Figure~\ref{f5}(b)), then immediately the circular ribbon and a remote brightening in the west start to strengthen. This initial remote brightening region is linked by the mildly hot loops to the southwest portion of the dome containing the NP, as aforementioned.

Based on these analyses, we propose that the present event could be triggered by magnetic reconnection between mildly twisted SA1 and SA2 \edit1{in agreement with the tether-cutting scenario \citep{moore01}, due to shearing and converging flows} that drive their inner two footpoints FP2 and FP3 (cf. Figures~\ref{f2}(c) and \ref{f5}(a)). This reconnection produces a more twisted \edit2{filament or FR}, which could become unstable attributed to the torus instability mainly acting on its southern portion (same as the region of SA2; see Figure~\ref{f3}(c)). This may account for the fact that the ejective motion along the southern slit S3 is more prominent than that along the northern slit S1 (cf. Figures~\ref{f4}(b) and (d)). As the southern cusped portion of the newly formed FR reaches and pushes against the NP, reconnection could then be induced at the NP. Accelerated electrons can precipitate along the fan to generate the circular ribbon, and also propagate predominantly along the high-$Q$ surfaces (Figure~\ref{f3}(d)) to cause the initial remote brightening in the west.

Stage II (the second row of Figure~\ref{f5}; after \sm05:11:32~UT): This phase of the event is accompanied with an impulsive, significant enhancement of HXR emissions and an accelerated eruption of the FR (Figures~\ref{f4}(a) and (c)). It is striking that after \sm05:12:08~UT (the vertical dashed line in Figure~\ref{f4}), the southern end of the FR is apparently detached from the surface (seen as a ``jump'' in Figure~\ref{f4}(c)) with its northern end still anchored, and hence the whole FR undergoes a whipping-like asymmetric eruption (see the animation; \citealt{liur09}); in the meantime, the remote brightening regions extend swiftly from the initial location mainly towards southeast along an arc reaching a scale of \sm400\arcsec\ (Figures~\ref{f5}(d)--(f)). The speed of the erupting FR is estimated to be about 660~\kms\ (Figure~\ref{f4}(c)), similar to the plane-of-sky speed of the associated CME (682~\kms) as recorded in the CDAW CME catalog \citep{yashiro04}. Noticeably, CME-associated coronal dimmings, a signature of sudden density depletion \citep[e.g.,][]{thompson00,harrison2000,harrison03}, are developed nearly cotemporal and cospatial with the extending remote brightenings (Figures~\ref{f5}(f)), and later (from \sm05:21~UT) also around the northern leg of the erupting FR (see the animation).

We speculate that because of the asymmetric strapping fields, the southern half of the formed FR might more easily break away from the surface and tear open the dome, leading to the whipping-like eruption and the ensued partial halo CME \citep{liu10,liu15}. This outward ejection may be able to sequentially disrupt and open the large-scale outer spine-like loops (cyan fields in Figures~\ref{f3}(d) and (e)) from north to south, causing rapidly moving remote brightenings at their footpoints \citep[e.g.,][]{liu06}. \edit2{Plasma leaves the corona and is injected (becoming part of the CME) along field lines} that are opened upward, causing coronal dimmings around the same locations of remote brightenings and subsequently also around the northern end of the FR. It is worth surmising that those disrupted outer spine-like loops may not need to be completely opened, as similar evolving coronal dimmings were also found in confined circular-ribbon flares without a CME \citep[e.g.,][]{zhang19}.

Finally, we examine the flare-related magnetic restructuring, as it can provide hints on the preflare magnetic field structure \citep{wang15}. Here we limit our investigation to the vector magnetic field changes on the surface. In Figure~\ref{f6}, we compare locations of rapid and permanent changes of \edit2{transverse} field $B_t$ with the essential preflare structures SA1 and SA2. The results show that in the red (blue) region, the mean \edit2{transverse} field increases (decreases) dramatically by \sm80\% (67\%) after the flare, and that the compact, enhanced $B_t$ region is spatially correlated with the strong shearing/converging flow region (cf. Figure~\ref{f2}(c)) right between the inner two footpoints FP2/FP3 of SA1 and SA2 (Figure~\ref{f5}(a)). This strongly indicates that SA1 and SA2 reconnect in this event, and the newly created shorter, low-lying loops connecting FP2-FP3 contribute to the strengthened $B_t$ in the small region between FP2 and FP3 along the PIL1 \citep[e.g.,][]{liu12,liu13}. The weakened $B_t$ region lies in the sunspot penumbrae swept by the later developed flare ribbon initially recognized as the footpoint FP2 (Figure~\ref{f5}(a)), and can be \edit2{the result of fields becoming} more vertical at those locations \citep[e.g.,][]{liu05,wang12b}.

\section{SUMMARY AND DISCUSSION}\label{summary}
In this paper, we have presented analyses of the 2013 November 10 X1.1 flare/CME event spawned by NOAA AR 11890. This event caught our attention because this type of \textit{eruptive} circular-ribbon flares are less studied, and most importantly, the entire evolution from the formation of the eruptive FR to its asymmetric eruption through the dome is clearly observed. Furthermore, unlike other circular-ribbon flares, the present event produces fast evolving, significantly extended remote brightenings. Taking advantage of multiwavelength observations and a NFFF model, we shed light on the event triggering and progression mechanisms based on investigation of magnetic structure and dynamics. Our major findings and interpretations are summarized as follows.

\begin{enumerate} 
\item In the preflare state, the trailing negative fields of this AR together with intruding compact positive fields in the center form a dome-like structure in EUV, with the coronal NP located in its southwest portion. Under the dome, mildly twisted SA1 and SA2 \edit2{lie close to} each other along the main PIL1. Around their inner two footpoints FP2 and FP3, pronounced shearing and converging flows are present. Also, SA2 approaching the NP could be in the state of torus-unstable regime. In large scale, outer spine-like loops that envelope the dome constitute a ``magnetic curtain'', hanging above and rooted in the extended arc-like remote brightening regions. Remarkably, mildly hot outer spine-line loops apparently link the southwest portion of the dome and the western initial remote brightening region, materializing part of the magnetic curtain with the highest $Q$.

\item This flare/CME event is composed of two stages which are visible as \edit2{a series of minor peaks} and one major peak in HXRs. In stage I, SA1 and SA2 reconnect due to shearing and converging flows in their inner footpoint region, triggering the event. This is also evidenced by the substantially and permanently enhanced $B_t$ on the photosphere between FP2 and FP3 after the flare. The southern portion of the formed FR close to the NP subsequently rises upward due to the torus instability. When the FR reaches the fan surface, \edit2{NP reconnection leads to the formation of} the circular chromospheric ribbon. Energy also flows outward primarily along the high-$Q$ surfaces to ignite the initial remote brightening in the west. In stage II, the southern end of the FR detaches from the surface and rapidly erupts (at 660~\kms) in a whipping fashion to become the partial halo CME. This opens the dome and the outer spine-like loops sequentially, resulting in fast-moving remote brightenings spanning a large range (\sm400\arcsec) at the footpoints of the magnetic curtain and the formation of the cospatial coronal dimmings. Correspondingly, much stronger HXR and ribbon emissions are produced in this phase than in stage I.
\end{enumerate}

It can be noted that the NFFF model reasonably predicts all the topological structures as suggested by observations. Further MHD simulation using such a preflare NFFF model as initial condition has potential to disclose more detailed interaction process between the erupting FR and outer overlying fields \citep{prasad18,prasad19,nayak19}.

\acknowledgments
We thank the teams of \sdo\ and \hsi\ for the observational data products, \edit1{and the referee for valuable comments that helped us improve the paper}. C.L., J.L., and H.W. were supported by NASA grants 80NSSC17K0016, 80NSSC18K0673, and 80NSSC18K1705, and by NSF grants AGS-1821294, AGS-1927578, and AGS-1954734. A.P. acknowledges partial support by NASA grant\\80NSSC17K0016 and NSF awards AGS-1650854 and AGS-1954503. This work utilizes the NFFF and DAVE4VM codes written and developed by Qiang Hu and Peter W. Schuck, respectively.

\vspace{5mm}
\facilities{\sdo, \hsi}

\begin{figure*}
\epsscale{1.17}
\plotone{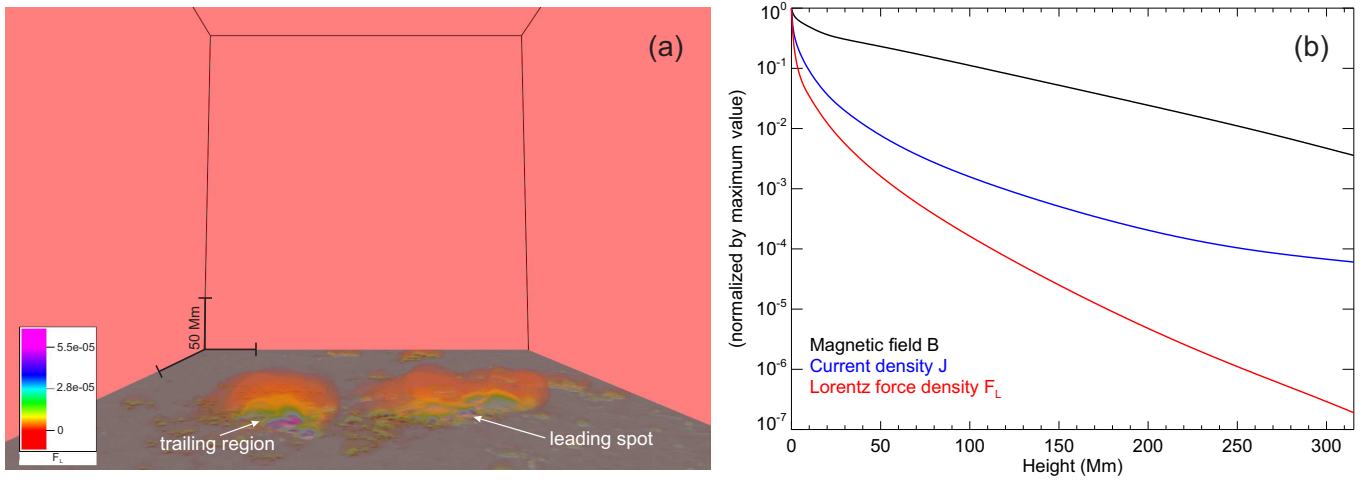}
\caption{Spatial distribution of physical parameters in the NFFF model. (a) Direct volume rendering of Lorentz force density $F_L$. The unit of color bar is dynes~cm$^{-3}$. \edit2{The black lines are bounds of the computation domain, which has a size of about 316~$\times$~467~$\times$~316~Mm$^3$}. (b) Variation of magnetic field strength $B$, current density $J$, and $F_L$ with height. The values are averaged over the circular-ribbon flare area (in the trailing region) at each height and normalized by the maxima. \label{f7}} \end{figure*}

\appendix
\section{The Non-Force-Free Field Extrapolation}
The rationale behind the NFFF extrapolation lies in the dimensional analysis as follows \citep{mitra2018}:
\begin{equation}
\label{forcebalance}
\frac{|\mathbf{j}\times \mathbf{B}|}{|\rho \frac{d\mathbf{v}}{dt}|}\sim\frac{B^2}{L}\frac{t}{\rho v}\sim\frac{B^2}{\rho v^2}\sim\frac{B^2}{\rho {v_{th}}^2}\frac{v_{th}^2}{v^2}\sim\frac{1}{\beta}\frac{v_{th}^2}{v^2}
\end{equation}
where ${\mathbf{v_{th}}}$ and $\mathbf{j}=\nabla\times
\mathbf{B}$ are respectively the thermal velocity and the volume current density. It is generally accepted that the photospheric flow speed is $\sim$1 km s$^{-1}$ \citep{Vekstein2016,Khlystova2017}. The thermal speed of the photospheric plasma can be straightforwardly found to be also $\sim$1 km s$^{-1}$. Thus, the fact of ${v_{th}}\sim v$ on the photosphere leads to

\begin{equation}
\frac{|\mathbf{j}\times \mathbf{B}|}{|\rho \frac{d\mathbf{v}}{dt}|}\sim\frac{1}{\beta}~~.
\label{lorentz}
\end{equation}
Here, $\beta$ represents the ratio of thermal to magnetic pressure, and with an equipartition of kinetic and magnetic energies, it can be on the order of unity on the photosphere. In such conditions, Equation (\ref{lorentz}) then becomes
\begin{equation}
|\mathbf{j}\times\mathbf{B}|\approx|\rho\frac{d\mathbf{v}}{dt}|~~,
\end{equation}
showing the importance of the Lorentz force when $\beta\approx 1$.

The coronal magnetic field of the active region is obtained by using the numerical non-force free extrapolation code developed by 
\citet{hu08a,hu08b,hu10}. The code is based on the principle of minimum dissipation rate (MDR) \citep{montgomery88,dasgupta98,bhattacharyya04,bhattacharyya07}
---used extensively in literature to obtain dissipative relaxed states. 
The magnetic field $\mathbf{B}$ and the fluid vorticity $\boldsymbol{\omega}$ 
of an MDR relaxed state are given by \citep{bhattacharyya04,bhattacharyya07}
\begin{align}
\nabla\times(\nabla\times\mathbf{B})+a_1\nabla\times\mathbf{B}+b_1\mathbf{B}&=\nabla \psi\label{e:bnff}\\
\nabla\times(\nabla\times\boldsymbol{\omega})+a_2\nabla\times\boldsymbol{\omega}+b_2\boldsymbol{\omega}&=\nabla \chi
\end{align}
where $a_1$, $a_2$, $b_1$, $b_2$, are constants depending on the parameters of the system, and $\psi$ and $\chi$ are scalar functions satisfying the Laplace's equation. The above is the double-curl Beltrami equation representing a steady-state of two-fluid plasma \citep{mahajan98}.
Focusing only on the magnetic field, and taking a curl of Equation \eqref{e:bnff}, we obtain the following equation \citep{hu08b}
\begin{equation}
\nabla\times(\nabla\times(\nabla\times\mathbf{B}))+a_1\nabla\times(\nabla\times\mathbf{B})+b_1\nabla\times\mathbf{B}=0.\label{e:bnff3}
\end{equation}
An exact solution of Equation \eqref{e:bnff3} can be constructed using the Chandrasekhar-Kendall (CK) eigenfunctions \citep{chandrasekhar57} which also satisfy the linear force-free equation. Thus, the $\mathbf{B}$ satisfying Equation \eqref{e:bnff3} is written as \citep{hu08b}:
\begin{equation}
\mathbf{B} = \mathbf{B_1}+\mathbf{B_2}+\mathbf{B_3}; \quad \nabla\times\mathbf{B_i}=\alpha_i\mathbf{B_i}
\label{e:b123}
\end{equation}
where $\alpha_i$ are constants and $i=1,2,3$. The requirement of $\mathbf{B}$ in Equation \eqref{e:b123} to satisfy Equation \eqref{e:bnff3} with distinct values of $\alpha_i$ requires one of them to be zero. Here we arbitrarily set $\alpha_2=0$, making the corresponding magnetic field $\mathbf{B_2}$ to be potential. This further implies  $a_1=-(\alpha_1+\alpha_3)$ and $b_1=\alpha_1 \alpha_3$. 

For completeness, in the following we summarize salient features of the algorithm which solves Equation \eqref{e:bnff3}.  Combining Equations  \eqref{e:bnff3} and \eqref{e:b123}, we obtain

\begin{align}
\begin{pmatrix}
\mathbf{B_1}\\
\mathbf{B_2}\\
\mathbf{B_3}\\
\end{pmatrix}
=\mathcal{V}^{-1}
\begin{pmatrix}
\mathbf{B}\\
\nabla\times\mathbf{B}\\
\nabla\times\nabla\times\mathbf{B}\\
\end{pmatrix}
\label{e:vmat}
\end{align}
where the matrix $\mathcal{V}$ is a Vandermonde matrix comprising of elements $\alpha^{i-1}_j$ for $i, j = 1, 2, 3$ \citep{hu08a}.
Thus each linear force-free field (LFFF) ($\mathbf{B_i}$) can be obtained by assuming a value for the $\alpha$ parameter and using the normal boundary condition obtained from observations through a standard LFFF solver \citep{alissandrakis81}. With $\alpha_2$ = 0, an optimal pair of ($\alpha_1$ , $\alpha_3$) parameters is obtained by a trial-and-error process which finds the
 pair that minimizes the average deviation between the observed ($\mathbf{B}_t$) and the calculated ($\mathbf{b}_t$) transverse field, as indicated by the following metric \citep{hu08a,hu08b}:
\begin{equation}
E_n =\sum_{i=1}^M |\mathbf{B}_{t,i}-\mathbf{b}_{t,i}|/\sum_{i=1}^M |\mathbf{B}_{t,i}|
\end{equation}
where $M=N^2$, represents the total number of grids points on the transverse plane.
However, the right-hand side of Equation \eqref{e:vmat} can provide the boundary conditions (vertical components) for each sub-field, given the $\alpha$ parameters only if vector magnetograms are available at two or more layers. This is inevitable since the calculation involves the evaluation of the second-order derivative, $(\nabla\times\nabla\times\mathbf{B})_z=-\nabla^2 B_z$, at $z=0$. In order to work with the available single layer vector magnetograms, an algorithm was devised by \citet{hu10}, which involved additional iterations to successively correct the potential subfield $\mathbf{B_2}$. Starting with an initial guess, the simplest being $\mathbf{B_2}=0$, the system of Equation \eqref{e:vmat} is reduced to 2nd-order which allows for the determination of boundary conditions for $\mathbf{B_1}$ and $\mathbf{B_3}$, using the trial-and-error process as described above. If the resulting minimum $E_n$ value is not satisfactory, then a corrector potential field to $\mathbf{B_2}$ is derived from the difference transverse field, i.e., $\mathbf{B}_t-\mathbf{b}_t$, and added to the previous $\mathbf{B_2}$, in anticipation of improved match between the transverse fields, as measured by $E_n$. The algorithm relies on the implementation of fast calculations of the LFFFs including the potential field. For the present case, the best-fit $(\alpha_1 , \alpha_3)$ values obtained after 4000 iterations are $(0.0041, -0.0041)$~pixel$^{-1}$, which corresponds to an $E_n =0.30$. This residual error is similar to those in previous NFFF extrapolations \citep[e.g.,][]{prasad18,mitra2018}.

In Figure~\ref{f7}, we show the height-dependent variation of several physical quantities. It can be seen that most of the Lorentz force $F_L$ is concentrated very close to the photosphere (Figure~\ref{f7}(a)) and it decays very sharply with the height (Figure~\ref{f7}(b)). As we approach the corona, its strength falls more than four magnitudes of its photospheric value; in comparison, the current shows a much slower decay. This suggests that with increasing height, the current becomes more and more field-aligned so the coronal loops are in a very close to force-free equilibrium state; however, their footpoints are forced, which is in agreement with the generally accepted picture \citep{prasad18}.


\begin{thebibliography}{}
\expandafter\ifx\csname natexlab\endcsname\relax\def\natexlab#1{#1}\fi
\providecommand{\url}[1]{\href{#1}{#1}}

\bibitem[{{Alissandrakis}(1981)}]{alissandrakis81}
{Alissandrakis}, C.~E. 1981, \aap, 100, 197

\bibitem[{{Bateman}(1978)}]{bateman78}
{Bateman}, G. 1978, {MHD Instabilities} (Cambridge, MA: MIT Press)

\bibitem[{{Berger} \& {Prior}(2006)}]{berger06}
{Berger}, M.~A., \& {Prior}, C. 2006, Journal of Physics A Mathematical
  General, 39, 8321

\bibitem[{{Bhattacharyya} \& {Janaki}(2004)}]{bhattacharyya04}
{Bhattacharyya}, R., \& {Janaki}, M.~S. 2004, Physics of Plasmas, 11, 5615

\bibitem[{{Bhattacharyya} {et~al.}(2007){Bhattacharyya}, {Janaki}, {Dasgupta},
  \& {Zank}}]{bhattacharyya07}
{Bhattacharyya}, R., {Janaki}, M.~S., {Dasgupta}, B., \& {Zank}, G.~P. 2007,
  \solphys, 240, 63

\bibitem[{{Borrero} {et~al.}(2011){Borrero}, {Tomczyk}, {Kubo},
  {Socas-Navarro}, {Schou}, {Couvidat}, \& {Bogart}}]{borrero11}
{Borrero}, J.~M., {Tomczyk}, S., {Kubo}, M., {et~al.} 2011, \solphys, 273, 267

\bibitem[{{Chandra} {et~al.}(2017){Chandra}, {Mandrini}, {Schmieder}, {Joshi},
  {Cristiani}, {Cremades}, {Pariat}, {Nuevo}, {Srivastava}, \&
  {Uddin}}]{chandra17}
{Chandra}, R., {Mandrini}, C.~H., {Schmieder}, B., {et~al.} 2017, \aap, 598,
  A41

\bibitem[{{Chandrasekhar} \& {Kendall}(1957)}]{chandrasekhar57}
{Chandrasekhar}, S., \& {Kendall}, P.~C. 1957, \apj, 126, 457

\bibitem[{{Dasgupta} {et~al.}(1998){Dasgupta}, {Dasgupta}, {Janaki},
  {Watanabe}, \& {Sato}}]{dasgupta98}
{Dasgupta}, B., {Dasgupta}, P., {Janaki}, M.~S., {Watanabe}, T., \& {Sato}, T.
  1998, Physical Review Letters, 81, 3144

\bibitem[{{D{\'e}moulin} \& {Aulanier}(2010)}]{demoulin10}
{D{\'e}moulin}, P., \& {Aulanier}, G. 2010, \apj, 718, 1388

\bibitem[{{D{\'e}moulin} {et~al.}(1996){D{\'e}moulin}, {Henoux}, {Priest}, \&
  {Mandrini}}]{demoulin96}
{D{\'e}moulin}, P., {Henoux}, J.~C., {Priest}, E.~R., \& {Mandrini}, C.~H.
  1996, \aap, 308, 643

\bibitem[{{Deng} {et~al.}(2011){Deng}, {Liu}, {Prasad Choudhary}, \&
  {Wang}}]{deng11}
{Deng}, N., {Liu}, C., {Prasad Choudhary}, D., \& {Wang}, H. 2011, \apjl, 733,
  L14

\bibitem[{{Deng} {et~al.}(2006){Deng}, {Xu}, {Yang}, {Cao}, {Liu}, {Rimmele},
  {Wang}, \& {Denker}}]{deng06}
{Deng}, N., {Xu}, Y., {Yang}, G., {et~al.} 2006, \apj, 644, 1278

\bibitem[{{Galsgaard} \& {Nordlund}(1997)}]{galsgaard97}
{Galsgaard}, K., \& {Nordlund}, {\AA}. 1997, \jgr, 102, 231

\bibitem[{{Galsgaard} {et~al.}(2003){Galsgaard}, {Priest}, \&
  {Titov}}]{galsgaard03}
{Galsgaard}, K., {Priest}, E.~R., \& {Titov}, V.~S. 2003, Journal of
  Geophysical Research (Space Physics), 108, 1042

\bibitem[{{Harrison} {et~al.}(2003){Harrison}, {Bryans}, {Simnett}, \&
  {Lyons}}]{harrison03}
{Harrison}, R.~A., {Bryans}, P., {Simnett}, G.~M., \& {Lyons}, M. 2003, \aap,
  400, 1071

\bibitem[{{Harrison} \& {Lyons}(2000)}]{harrison2000}
{Harrison}, R.~A., \& {Lyons}, M. 2000, \aap, 358, 1097

\bibitem[{{Hoeksema} {et~al.}(2014){Hoeksema}, {Liu}, {Hayashi}, {Sun},
  {Schou}, {Couvidat}, {Norton}, {Bobra}, {Centeno}, {Leka}, {Barnes}, \&
  {Turmon}}]{hoeksema14}
{Hoeksema}, J.~T., {Liu}, Y., {Hayashi}, K., {et~al.} 2014, \solphys, 289, 3483

\bibitem[{{Hu} \& {Dasgupta}(2008)}]{hu08a}
{Hu}, Q., \& {Dasgupta}, B. 2008, \solphys, 247, 87

\bibitem[{{Hu} {et~al.}(2008){Hu}, {Dasgupta}, {Choudhary}, \&
  {B{\"u}chner}}]{hu08b}
{Hu}, Q., {Dasgupta}, B., {Choudhary}, D.~P., \& {B{\"u}chner}, J. 2008, \apj,
  679, 848

\bibitem[{{Hu} {et~al.}(2010){Hu}, {Dasgupta}, {Derosa}, {B{\"u}chner}, \&
  {Gary}}]{hu10}
{Hu}, Q., {Dasgupta}, B., {Derosa}, M.~L., {B{\"u}chner}, J., \& {Gary}, G.~A.
  2010, Journal of Atmospheric and Solar-Terrestrial Physics, 72, 219

\bibitem[{{Janvier} {et~al.}(2013){Janvier}, {Aulanier}, {Pariat}, \&
  {D{\'e}moulin}}]{janvier13}
{Janvier}, M., {Aulanier}, G., {Pariat}, E., \& {D{\'e}moulin}, P. 2013, \aap,
  555, A77

\bibitem[{{Jiang} {et~al.}(2014){Jiang}, {Wu}, {Feng}, \& {Hu}}]{jiang14}
{Jiang}, C., {Wu}, S.~T., {Feng}, X., \& {Hu}, Q. 2014, \apj, 780, 55

\bibitem[{{Joshi} {et~al.}(2015){Joshi}, {Liu}, {Sun}, {Wang}, {Magara}, \&
  {Moon}}]{joshi15}
{Joshi}, N.~C., {Liu}, C., {Sun}, X., {et~al.} 2015, \apj, 812, 50

\bibitem[{{Khlystova} \& {Toriumi}(2017)}]{Khlystova2017}
{Khlystova}, A., \& {Toriumi}, S. 2017, \apj, 839, 63

\bibitem[{{Kliem} \& {T{\"o}r{\"o}k}(2006)}]{kliem06}
{Kliem}, B., \& {T{\"o}r{\"o}k}, T. 2006, Phys. Rev. Lett., 96, 255002

\bibitem[{{Kopp} \& {Pneuman}(1976)}]{kopp76}
{Kopp}, R.~A., \& {Pneuman}, G.~W. 1976, \solphys, 50, 85

\bibitem[{{Lau} \& {Finn}(1990)}]{lau90}
{Lau}, Y.-T., \& {Finn}, J.~M. 1990, \apj, 350, 672

\bibitem[{{Lee} {et~al.}(2016){Lee}, {Liu}, {Jing}, \& {Chae}}]{lee16a}
{Lee}, J., {Liu}, C., {Jing}, J., \& {Chae}, J. 2016, \apj, 829, L1

\bibitem[{{Lemen} {et~al.}(2012){Lemen}, {Title}, {Akin}, {Boerner}, {Chou},
  {Drake}, {Duncan}, {Edwards}, {Friedlaender}, {Heyman}, {Hurlburt}, {Katz},
  {Kushner}, {Levay}, {Lindgren}, {Mathur}, {McFeaters}, {Mitchell}, {Rehse},
  {Schrijver}, {Springer}, {Stern}, {Tarbell}, {Wuelser}, {Wolfson}, {Yanari},
  {Bookbinder}, {Cheimets}, {Caldwell}, {Deluca}, {Gates}, {Golub}, {Park},
  {Podgorski}, {Bush}, {Scherrer}, {Gummin}, {Smith}, {Auker}, {Jerram},
  {Pool}, {Soufli}, {Windt}, {Beardsley}, {Clapp}, {Lang}, \&
  {Waltham}}]{lemen12}
{Lemen}, J.~R., {Title}, A.~M., {Akin}, D.~J., {et~al.} 2012, \solphys, 275, 17

\bibitem[{{Lin} {et~al.}(2002){Lin}, {Dennis}, {Hurford}, {Smith}, {Zehnder},
  {Harvey}, {Curtis}, {Pankow}, {Turin}, {Bester}, {Csillaghy}, {Lewis},
  {Madden}, {van Beek}, {Appleby}, {Raudorf}, {McTiernan}, {Ramaty}, {Schmahl},
  {Schwartz}, {Krucker}, {Abiad}, {Quinn}, {Berg}, {Hashii}, {Sterling},
  {Jackson}, {Pratt}, {Campbell}, {Malone}, {Landis}, {Barrington-Leigh},
  {Slassi-Sennou}, {Cork}, {Clark}, {Amato}, {Orwig}, {Boyle}, {Banks},
  {Shirey}, {Tolbert}, {Zarro}, {Snow}, {Thomsen}, {Henneck}, {McHedlishvili},
  {Ming}, {Fivian}, {Jordan}, {Wanner}, {Crubb}, {Preble}, {Matranga}, {Benz},
  {Hudson}, {Canfield}, {Holman}, {Crannell}, {Kosugi}, {Emslie}, {Vilmer},
  {Brown}, {Johns-Krull}, {Aschwanden}, {Metcalf}, \& {Conway}}]{lin02}
{Lin}, R.~P., {Dennis}, B.~R., {Hurford}, G.~J., {et~al.} 2002, \solphys, 210,
  3

\bibitem[{{Liu} {et~al.}(2013{\natexlab{a}}){Liu}, {Deng}, {Lee}, {Wiegelmann},
  {Moore}, \& {Wang}}]{liu13}
{Liu}, C., {Deng}, N., {Lee}, J., {et~al.} 2013{\natexlab{a}}, \apjl, 778, L36

\bibitem[{{Liu} {et~al.}(2005){Liu}, {Deng}, {Liu}, {Falconer}, {Goode},
  {Denker}, \& {Wang}}]{liu05}
{Liu}, C., {Deng}, N., {Liu}, Y., {et~al.} 2005, \apj, 622, 722

\bibitem[{{Liu} {et~al.}(2006){Liu}, {Lee}, {Deng}, {Gary}, \& {Wang}}]{liu06}
{Liu}, C., {Lee}, J., {Deng}, N., {Gary}, D.~E., \& {Wang}, H. 2006, \apj, 642,
  1205

\bibitem[{{Liu} {et~al.}(2010){Liu}, {Lee}, {Jing}, {Liu}, {Deng}, \&
  {Wang}}]{liu10}
{Liu}, C., {Lee}, J., {Jing}, J., {et~al.} 2010, \apjl, 721, L193

\bibitem[{{Liu} {et~al.}(2019){Liu}, {Lee}, \& {Wang}}]{liuc19}
{Liu}, C., {Lee}, J., \& {Wang}, H. 2019, \apj, 883, 47

\bibitem[{{Liu} {et~al.}(2012){Liu}, {Deng}, {Liu}, {Lee}, {Wiegelmann},
  {Jing}, {Xu}, {Wang}, \& {Wang}}]{liu12}
{Liu}, C., {Deng}, N., {Liu}, R., {et~al.} 2012, \apjl, 745, L4

\bibitem[{{Liu} {et~al.}(2015){Liu}, {Deng}, {Liu}, {Lee}, {Pariat},
  {Wiegelmann}, {Liu}, {Kleint}, \& {Wang}}]{liu15}
---. 2015, \apjl, 812, L19

\bibitem[{{Liu} {et~al.}(2009){Liu}, {Alexander}, \& {Gilbert}}]{liur09}
{Liu}, R., {Alexander}, D., \& {Gilbert}, H.~R. 2009, \apj, 691, 1079

\bibitem[{{Liu} {et~al.}(2016){Liu}, {Kliem}, {Titov}, {Chen}, {Wang}, {Wang},
  {Liu}, {Xu}, \& {Wiegelmann}}]{liur16}
{Liu}, R., {Kliem}, B., {Titov}, V.~S., {et~al.} 2016, \apj, 818, 148

\bibitem[{{Liu}(2008)}]{liuy08}
{Liu}, Y. 2008, \apjl, 679, L151

\bibitem[{{Liu} {et~al.}(2013{\natexlab{b}}){Liu}, {Zhao}, \&
  {Schuck}}]{liuy13}
{Liu}, Y., {Zhao}, J., \& {Schuck}, P.~W. 2013{\natexlab{b}}, \solphys, 287,
  279

\bibitem[{{L{\'o}pez Fuentes} {et~al.}(2018){L{\'o}pez Fuentes}, {Mandrini},
  {Poisson}, {D{\'e}moulin}, {Cristiani}, {L{\'o}pez}, \& {Luoni}}]{fuentes18}
{L{\'o}pez Fuentes}, M., {Mandrini}, C.~H., {Poisson}, M., {et~al.} 2018,
  \solphys, 293, 166

\bibitem[{{Mahajan} \& {Yoshida}(1998)}]{mahajan98}
{Mahajan}, S.~M., \& {Yoshida}, Z. 1998, Physical Review Letters, 81, 4863

\bibitem[{{Masson} {et~al.}(2009){Masson}, {Pariat}, {Aulanier}, \&
  {Schrijver}}]{masson09}
{Masson}, S., {Pariat}, E., {Aulanier}, G., \& {Schrijver}, C.~J. 2009, \apj,
  700, 559

\bibitem[{{Masson} {et~al.}(2017){Masson}, {Pariat}, {Valori}, {Deng}, {Liu},
  {Wang}, \& {Reid}}]{masson17}
{Masson}, S., {Pariat}, {\'E}., {Valori}, G., {et~al.} 2017, \aap, 604, A76

\bibitem[{{Mitra} {et~al.}(2018){Mitra}, {Joshi}, {Prasad}, {Veronig}, \&
  {Bhattacharyya}}]{mitra2018}
{Mitra}, P.~K., {Joshi}, B., {Prasad}, A., {Veronig}, A.~M., \&
  {Bhattacharyya}, R. 2018, \apj, 869, 69

\bibitem[{{Montgomery} \& {Phillips}(1988)}]{montgomery88}
{Montgomery}, D., \& {Phillips}, L. 1988, \pra, 38, 2953

\bibitem[{{Moore} {et~al.}(2001){Moore}, {Sterling}, {Hudson}, \&
  {Lemen}}]{moore01}
{Moore}, R.~L., {Sterling}, A.~C., {Hudson}, H.~S., \& {Lemen}, J.~R. 2001,
  \apj, 552, 833

\bibitem[{{Nayak} {et~al.}(2019){Nayak}, {Bhattacharyya}, {Prasad}, {Hu},
  {Kumar}, \& {Joshi}}]{nayak19}
{Nayak}, S.~S., {Bhattacharyya}, R., {Prasad}, A., {et~al.} 2019, \apj, 875, 10

\bibitem[{{Pesnell} {et~al.}(2012){Pesnell}, {Thompson}, \&
  {Chamberlin}}]{pesnell12}
{Pesnell}, W.~D., {Thompson}, B.~J., \& {Chamberlin}, P.~C. 2012, \solphys,
  275, 3

\bibitem[{{Pontin} {et~al.}(2013){Pontin}, {Priest}, \& {Galsgaard}}]{pontin13}
{Pontin}, D.~I., {Priest}, E.~R., \& {Galsgaard}, K. 2013, \apj, 774, 154

\bibitem[{{Prasad} {et~al.}(2018){Prasad}, {Bhattacharyya}, {Hu}, {Kumar}, \&
  {Nayak}}]{prasad18}
{Prasad}, A., {Bhattacharyya}, R., {Hu}, Q., {Kumar}, S., \& {Nayak}, S.~S.
  2018, \apj, 860, 96

\bibitem[{{Prasad} {et~al.}(2017){Prasad}, {Bhattacharyya}, \&
  {Kumar}}]{prasad17}
{Prasad}, A., {Bhattacharyya}, R., \& {Kumar}, S. 2017, \apj, 840, 37

\bibitem[{{Prasad} {et~al.}(2019){Prasad}, {Dissauer}, {Hu}, {Bhattacharyya},
  {Veronig}, {Kumar}, \& {Joshi}}]{prasad19}
{Prasad}, A., {Dissauer}, K., {Hu}, Q., {et~al.} 2019, in American Astronomical
  Society Meeting Abstracts, Vol.~51, American Astronomical Society Meeting
  Abstracts \#234, 310.04

\bibitem[{{Reid} {et~al.}(2012){Reid}, {Vilmer}, {Aulanier}, \&
  {Pariat}}]{reid12}
{Reid}, H.~A.~S., {Vilmer}, N., {Aulanier}, G., \& {Pariat}, E. 2012, \aap,
  547, A52

\bibitem[{{Schou} {et~al.}(2012){Schou}, {Scherrer}, {Bush}, {Wachter},
  {Couvidat}, {Rabello-Soares}, {Bogart}, {Hoeksema}, {Liu}, {Duvall}, {Akin},
  {Allard}, {Miles}, {Rairden}, {Shine}, {Tarbell}, {Title}, {Wolfson},
  {Elmore}, {Norton}, \& {Tomczyk}}]{schou12}
{Schou}, J., {Scherrer}, P.~H., {Bush}, R.~I., {et~al.} 2012, \solphys, 275,
  229

\bibitem[{{Schrijver}(2016)}]{schrijver16}
{Schrijver}, C.~J. 2016, \apj, 820, 103

\bibitem[{{Schuck}(2008)}]{schuck08}
{Schuck}, P.~W. 2008, \apj, 683, 1134

\bibitem[{{Song} \& {Tian}(2018)}]{song18}
{Song}, Y., \& {Tian}, H. 2018, \apj, 867, 159

\bibitem[{{Sun} {et~al.}(2013){Sun}, {Hoeksema}, {Liu}, {Aulanier}, {Su},
  {Hannah}, \& {Hock}}]{sun13}
{Sun}, X., {Hoeksema}, J.~T., {Liu}, Y., {et~al.} 2013, \apj, 778, 139

\bibitem[{{Sun} {et~al.}(2017){Sun}, {Hoeksema}, {Liu}, {Kazachenko}, \&
  {Chen}}]{sun17}
{Sun}, X., {Hoeksema}, J.~T., {Liu}, Y., {Kazachenko}, M., \& {Chen}, R. 2017,
  \apj, 839, 67

\bibitem[{{Thompson} {et~al.}(2000){Thompson}, {Cliver}, {Nitta},
  {Delann{\'e}e}, \& {Delaboudini{\`e}re}}]{thompson00}
{Thompson}, B.~J., {Cliver}, E.~W., {Nitta}, N., {Delann{\'e}e}, C., \&
  {Delaboudini{\`e}re}, J.-P. 2000, \grl, 27, 1431

\bibitem[{{Titov}(2007)}]{titov07}
{Titov}, V.~S. 2007, \apj, 660, 863

\bibitem[{{Titov} {et~al.}(2002){Titov}, {Hornig}, \& {D{\'e}moulin}}]{titov02}
{Titov}, V.~S., {Hornig}, G., \& {D{\'e}moulin}, P. 2002, Journal of
  Geophysical Research (Space Physics), 107, 1164

\bibitem[{{Toriumi} \& {Wang}(2019)}]{toriumi19}
{Toriumi}, S., \& {Wang}, H. 2019, Living Reviews in Solar Physics, 16, 3

\bibitem[{{T{\"o}r{\"o}k} {et~al.}(2009){T{\"o}r{\"o}k}, {Aulanier},
  {Schmieder}, {Reeves}, \& {Golub}}]{torok09}
{T{\"o}r{\"o}k}, T., {Aulanier}, G., {Schmieder}, B., {Reeves}, K.~K., \&
  {Golub}, L. 2009, \apj, 704, 485

\bibitem[{{T{\"o}r{\"o}k} \& {Kliem}(2007)}]{torok07}
{T{\"o}r{\"o}k}, T., \& {Kliem}, B. 2007, Astronomische Nachrichten, 328, 743

\bibitem[{{Vekstein}(2016)}]{Vekstein2016}
{Vekstein}, G. 2016, Journal of Plasma Physics, 82, 925820401

\bibitem[{{Wang} {et~al.}(2015){Wang}, {Cao}, {Liu}, {Xu}, {Liu}, {Zeng},
  {Chae}, \& {Ji}}]{wang15b}
{Wang}, H., {Cao}, W., {Liu}, C., {et~al.} 2015, Nature Communications, 6, 7008

\bibitem[{{Wang} {et~al.}(2012){Wang}, {Deng}, \& {Liu}}]{wang12b}
{Wang}, H., {Deng}, N., \& {Liu}, C. 2012, \apj, 748, 76

\bibitem[{{Wang} \& {Liu}(2012)}]{wang12}
{Wang}, H., \& {Liu}, C. 2012, \apj, 760, 101

\bibitem[{{Wang} \& {Liu}(2015)}]{wang15}
---. 2015, RAA, 15, 145

\bibitem[{{Welsch} {et~al.}(2009){Welsch}, {Li}, {Schuck}, \&
  {Fisher}}]{welsch09}
{Welsch}, B.~T., {Li}, Y., {Schuck}, P.~W., \& {Fisher}, G.~H. 2009, \apj, 705,
  821

\bibitem[{{Yang} {et~al.}(2004){Yang}, {Xu}, {Cao}, {Wang}, {Denker}, \&
  {Rimmele}}]{yang04}
{Yang}, G., {Xu}, Y., {Cao}, W., {et~al.} 2004, \apjl, 617, L151

\bibitem[{{Yang} {et~al.}(2015){Yang}, {Guo}, \& {Ding}}]{yang15}
{Yang}, K., {Guo}, Y., \& {Ding}, M.~D. 2015, \apj, 806, 171

\bibitem[{{Yashiro} {et~al.}(2004){Yashiro}, {Gopalswamy}, {Michalek},
  {St.~Cyr}, {Plunkett}, {Rich}, \& {Howard}}]{yashiro04}
{Yashiro}, S., {Gopalswamy}, N., {Michalek}, G., {et~al.} 2004, \jgr, 109,
  A07105

\bibitem[{{Zhang} {et~al.}(2015){Zhang}, {Ning}, {Guo}, {Zhou}, {Cheng}, {Ji},
  {Feng}, \& {Wiegelmann}}]{zhang15}
{Zhang}, Q.~M., {Ning}, Z.~J., {Guo}, Y., {et~al.} 2015, \apj, 805, 4

\bibitem[{{Zhang} \& {Zheng}(2020)}]{zhang19}
{Zhang}, Q.~M., \& {Zheng}, R.~S. 2020, A\&A, accepted, arXiv:1912.09618

\bibitem[{{Zhao} {et~al.}(2005){Zhao}, {Wang}, {Zhang}, \& {Xiao}}]{zhao05}
{Zhao}, H., {Wang}, J.-X., {Zhang}, J., \& {Xiao}, C.-J. 2005, \cjaa, 5, 443

\bibitem[{{Zhong} {et~al.}(2019){Zhong}, {Guo}, {Ding}, {Fang}, \&
  {Hao}}]{zhong19}
{Zhong}, Z., {Guo}, Y., {Ding}, M.~D., {Fang}, C., \& {Hao}, Q. 2019, \apj,
  871, 105

\end{thebibliography}
\end{document}